\documentclass[10pt]{article}
\usepackage{graphicx}
\usepackage{longtable}
\usepackage{times}
\usepackage{setspace}
\usepackage{dcolumn}
\usepackage{bm}
\usepackage{rotating}
\usepackage{hyperref}
\usepackage{epstopdf}
\usepackage{bm}
\usepackage{rotating}
\begin{document}
\begin{center}
    \textbf{Gravitational Redshift from Rotating Body Having Intense Magnetic Field}
\end{center}

\begin{center}
    $ Anuj \ Kumar \ Dubey^{1}$,  $ A. \ K. \ Sen^{2}$ 
\end{center}
\begin{center}
    \textit{Department of Physics, Assam University, Silchar-788011, Assam, India}
\end{center}
\begin{center}
    email: $ danuj67@gmail.com^{1}$, $ asokesen@yahoo.com^{2}$
\end{center}
\begin{abstract}
It is well known fact that gravitational field can alter the space-time structure and gravitational redshift is its one example. Electromagnetic field can also alter the space-time similar to gravitational field. So electromagnetic field can give rise to an additional effect on gravitational redshift. There are many objects in nature, like neutron stars, magnetars etc which have high amount of rotation and magnetic field. In the present paper we will derive the expression of gravitational redshift from rotating body having intense magnetic field by using the action function of the electromagnetic fields.
\end{abstract}
\section{Introduction}
General relativity is not only relativistic theory of gravitation proposed by Einstein, but it is the simplest theory that is consistent with experimental data. Gravitational redshift of light is one of the predictions of general relativity and also provides evidence for the validity of the principle of equivalence. Any relativistic theory of gravitation consistent with the principle of equivalence will predict a redshift.

 Gravitational redshift has been reported by most of the authors without consideration of rotation of a body. Neglecting the rotation, the geometry of space time can be described using the well-known spherically symmetric Schwarzschild's geometry and information on the ratio $ \frac{M}{r}$ of a compact object can be obtained from the gravitational redshift, where M and r are mass and radius respectively. Thus the redshifted angular frequency ($\omega '$) and the original angular frequency ($\omega $) of a photon in Schwarzschild geometry are related by the relation (page 268, of Landau and Lifshitz \cite{Landau})
\begin{equation}
\omega^{'}= \frac{\omega}{\sqrt{g_{tt}}}=\frac{\omega}{\sqrt{1-\frac{r_{g}}{r}}}
\end{equation}
where $r_g=(2GM/c^2)$ is the Schwarzschild radius.

Adams in 1925 has claimed first about the confirmation of the predicted gravitational redshift from the measurement of the apparent radial velocity of Sirius B \cite{Adams}. Pound and Rebka in 1959 were the first to experimentally verify the gravitational redshift from nuclear - resonance \cite{Pound and Rebka}. Pound and Snider in 1965 had performed an improved version of the experiment of Pound and Rebka, to measure the effect of gravity, making use of Mossbauer - Effect \cite{Pound and Snider}. Snider in 1972 has measured the redshift of the solar potassium absorption line at 7699 ${\AA}$ by using an atomic - beam resonance - scattering technique \cite{Snider}. Krisher et al. in 1993 had measured the gravitational redshift of Sun \cite{Krisher}. Nunez and Nowakowski in 2010 had obtained an expression for gravitational redshift factor of rotating body by using small perturbations to the Schwarzschild's geometry \cite{Nunez and Nowakowski}. Payandeh and Fathi in 2013 had obtained the gravitational redshift for a static spherically symmetric electrically charged object in Isotropic Reissner - Nordstrom Geometry \cite{Payandeh and Fathi}. Dubey and Sen in 2015 had obtained the expression for gravitational redshift from rotating body in Kerr Geometry. They also showed the rotation and the latitude dependence of gravitational redshift from a rotating body (such as pulsars) \cite{Dubey and Sen 2014}. The expression of Gravitational Redshift Factor ($\Re$) from rotating body in Kerr geometry was given as (Eqn. 69, Dubey and Sen \cite{Dubey and Sen 2014}):
\begin{equation}
\Re(\phi ,\theta)= \sqrt{{g_{tt}+ g_{\phi\phi}(\frac{d\phi}{c dt})^{2}} +2 g_{t \phi}(\frac{d\phi}{c dt})}
\end{equation}
Dubey and Sen in 2015 had obtained the expression for gravitational redshift from charged rotating body in Kerr - Newman Geometry. They showed that gravitational redshift increases as the electrostatic and magnetostatic charges increase, for a fixed value of latitude
at which light ray has been emitted. Gravitational redshift increases from pole to equatorial region (maximum at equator), for a given set of values for electrostatic and magnetostatic charge \cite{Dubey and Sen 2015}.

With this background in the present paper we will derive the expression of gravitational redshift from rotating body having intense magnetic field by using the action function of the electromagnetic fields.
\section{Geometry of Rotatating Body}
When rotation is taken into consideration, the covariant form of metric tensor for Kerr family (Kerr (1963) \cite{Kerr 1963}, Newman et al. (1965) \cite{Newman 1965}) in terms of Boyer-Lindquist coordinates with signature (+,-,-,-) is expressed as:
\begin{equation}
ds^{2} = g_{tt}c^{2}dt^{2}+ g_{rr} dr^{2} + g_{\theta\theta} d\theta^{2} +g_{\phi\phi} d\phi^{2} + 2 g_{t \phi} c dt d\phi
\end{equation}
where $g_{ij}$'s are non-zero components of Kerr family. \\
If we consider the three parameters: mass (M), rotation parameter (a) and charge (electric (Q) and / or magnetic (P)), then it is easy to include charge in the Non-zero components of $g_{ij}$ of Kerr metric, simply by replacing $(r_{g} r)$ with $(r_{g} r-Q^{2}-P^{2})$.
 \\ Non-zero components of $g_{ij}$ of Kerr-Newman metric are given as follows (page 261-262 of Carroll (2004) \cite{Carroll}):
\begin{equation}
 g_{tt} = (1-\frac{r_{g} r-Q^{2}-P^{2}}{\rho^{2}})
\end{equation}
\begin{equation}
g_{rr}=-\frac{\rho^{2}}{\Delta}
\end{equation}
 \begin{equation}
 g_{\theta\theta}=-{\rho^{2}}
 \end{equation}
 \begin{equation}
g_{\phi\phi}= -[r^{2}+a^{2}+\frac{(r_{g}r-Q^{2}-P^{2}) a^{2}sin^{2}\theta}{\rho^{2}}]sin^{2}\theta
 \end{equation}
\begin{equation}
g_{t\phi}=\frac{a sin^{2}\theta (r_{g}r-Q^{2}-P^{2})}{\rho^{2}}
\end{equation}
with
\begin{equation}
\rho^{2} = r^{2}+ a^{2}cos^{2}\theta
\end{equation}
and
\begin{equation}
 \Delta =r^{2}+ a^{2}-r_{g}r +Q^{2}+P^{2}
\end{equation}
where $r_g=(2GM/c^2)$ is the Schwarzschild radius. Q and P are electric and magnetic charges respectively and a $(= \frac{J}{Mc})$ is rotation parameter of the source.

If we replace $(r_{g} r-Q^{2}-P^{2})$ by $(r_{g} r-Q^{2})$ and further if we put rotation parameter of the source (a) equal to zero, then it reduces to Reissner - Nordstrom metric. Also  if we replace $(r_{g} r-Q^{2}-P^{2})$ by $(r_{g} r)$ then the Kerr-Newman metric reduces to Kerr metric and further if we put rotation parameter of the source (a) equal to zero then it reduces to Schwarzschild metric.
\section{The Action Function of the Electromagnetic Fields}
 The action function S for the whole system, consisting of an electromagnetic field as well as the particles located in it, must consist of three parts given (page 71, of Landau and Lifshitz [1]) as:
\begin{equation}
S=S_{f}+S_{m}+S_{mf}
\end{equation}
where $S_{m}$ is that part of the action which depends only on the properties of the particles that is just the action of free particles. The quantity $S_{mf}$ is that part of the action which depends on the interaction between the particles and the field. The quantity $S_{f}$ is that part of the action which depends only on the properties of the field itself, that is, $S_{f}$ the action for a field in the absence of charges.
Thus $S_{m}$, $S_{mf}$ and $S_{f}$ has the form as:
\begin{equation}
S_{m}=-\sum mc \int ds
\end{equation}
\begin{equation}
S_{mf}=-\sum \frac{e}{c} \int A_{k} dx^{k}
\end{equation}
\begin{equation}
S_{f}=-\frac{1}{16\pi c} \int F_{ik}F^{ik}d\Gamma
\end{equation}
Using equations (12), (13) and (14), the action equation (11) for field and particle can be written as:
\begin{equation}
S=-\sum mc \int ds-\sum  \int \frac{e}{c} A_{k} dx^{k}-\frac{1}{16\pi c} \int F_{ik}F^{ik}d\Gamma
\end{equation}
where, $d\Gamma = c dt \ dx \ dy \ dz$, is four dimensional volume element.
The potential $A_{k}$ is the potential at that point of space time at which the corresponding particle is located.\\
 The potential $(A_{k})$ and electromagnetic field tensor $(F_{ik})$, refer to actual field, that is, the external field plus the field produced by particles themselves; $A_{k}$ and $F_{ik}$ depends on the positions and velocities of the charges.\\
If we consider charges (e) to be distributed continuously in space. Then we can introduce the charge density $\rho$ such that $\rho dV$ is the charge contained in the volume dV. We can also replace the sum over the charges by an integral over the whole volume.\\
We can define the current four - vector as:
 \begin{equation}
j^{i}= \rho \frac{dx^{i}}{dt}
\end{equation}
The space component of this vector form of the current density vector can be written as:
\begin{equation}
\textbf{j}= \rho \textbf{v}
\end{equation}
where \textbf{v} is the velocity of the charge at given point. The time component of the current four - vector is $c\rho$. Thus $j^{i}$ can be written as:
\begin{equation}
j^{i}= (c\rho,\textbf{j})
\end{equation}
Now using equation (16), we can rewrite the second term of the action (as given by equation (15)) as:
\begin{equation}
-\sum  \int \frac{e}{c} A_{k} dx^{k} =-\frac{1}{c}\int \rho\frac{dx^{i}}{dt} A_{i} dV dt=-\frac{1}{c^{2}}\int  A_{i} j^{i}d\Gamma
\end{equation}
On substituting the value of $A_{i}j^{i}(=A_{0} j^{0}-\textbf{A.J} =c\rho\Phi- \textbf{A.J})$, the above equation (19) can be written as:
\begin{equation}
 -\frac{1}{c^{2}}\int  A_{i} j^{i}d\Gamma =-\frac{1}{c^{2}}\int  (\Phi c\rho) dV cdt +\frac{1}{c^{2}}\int \textbf{A} . \textbf{J} \ dV \ cdt
\end{equation}
The energy stored in the magnetic field in S.I. system is given (page 94 of Scheck \cite{Scheck} and page 213-214 of Jackson \cite{Jackson}) as:
\begin{equation}
 \frac{1}{2c}\int \textbf{A} . \textbf{J } dV  =  \frac{1}{2\mu_{0}}\int H^{2} dV
\end{equation}
In Gaussian system the above equation (21) of magnetic energy can be written as:
\begin{equation}
  \frac{1}{2c}\int \textbf{A} . \textbf{J } dV = \frac{1}{8\pi}\int H^{2} dV
\end{equation}
From above equation (22), we can write:
\begin{equation}
 \frac{1}{c^{2}} \int \textbf{A} . \textbf{J } dV = \frac{1}{4\pi \ c}\int H^{2} dV
\end{equation}
Using equation (23), we can write equation (20) as:
\begin{equation}
-\sum  \int \frac{e}{c} A_{k} dx^{k} =-\frac{1}{c^{2}}\int  (\Phi c\rho) dV cdt +\frac{1}{c^{2}}\int \textbf{A} . \textbf{J} \ dV \ cdt=-\frac{1}{c}\int  \Phi
\rho d\Gamma  +\frac{1}{4\pi c}\int H^{2}d\Gamma
\end{equation}
The above equation (24) is the second term of the action equation (as given by equation (15)).\\
We can write (page 73, of Landau and Lifshitz \cite{Landau}) as:
\begin{equation}
F_{ik}F^{ik} = 2 (H^{2}-E^{2})=  invarient
\end{equation}
Now using the above equation (25), we can write the third term of the action (as given by equation (15)) as:
\begin{equation}
-\frac{1}{16\pi \ c} \int F_{ik}F^{ik}d\Gamma = -\frac{1}{16\pi \ c} \int 2 \ (H^{2}-E^{2}) d\Gamma
\end{equation}
Using equations (24) and (25), the action equation (as given by equation (15)) can be rewritten as:
\begin{equation}
S=-\sum mc \int ds-\frac{1}{c}\int  \Phi \ \rho \ d\Gamma  +\frac{1}{4\pi \ c}\int H^{2} \ d\Gamma -\frac{1}{8\pi \ c} \int (H^{2}-E^{2}) \ d\Gamma
\end{equation}
After Simplification above equation (27) can we written as:
 \begin{equation}
S=-\sum mc \int ds-\frac{1}{c}\int  \Phi \ \rho \ d\Gamma   +\frac{1}{8\pi \ c} \int (H^{2}+E^{2}) \ d\Gamma
\end{equation}
\section{Gravitational Redshift from Rotating Body Having Intense Magnetic Field}
Apsel in 1978 - 1979 had discussed, that the motion of a particle in a combination of gravitational and electro and magneto static field can be determined from a variation principle of the form $\delta\int d\tau$ =0. The field and motion equations are actually identical to Maxwell - Einstein theory. The theory predicted that even in a field free region of space, electro and magneto static potentials can alter the phase of wave function and the life time of charged particle \cite{Apsel 1978, Apsel 1979}.\\
 The space time is a Riemannian space with metric $g_{ij}$, it is natural to assume that law of motion for a particle in a combination of gravitational and electromagnetic fields (\cite{Apsel 1978, Apsel 1979}) as:
\begin{equation}
 \delta\int d\tau=0
 \end{equation}
  where
 \begin{equation}
 d\tau=\frac{\sqrt{g_{ij}dx^{i}dx^{j}}+\frac{eA_{i}dx^{i}}{mc^{2}}}{c}
 \end{equation}
 The above equation (30) can be rewritten as:
 \begin{equation}
 c d\tau=ds=\sqrt{g_{ij}dx^{i}dx^{j}}+\frac{eA_{i}dx^{i}}{mc^{2}}
 \end{equation}
In case of rotating body, we can use the value of $ds^{2} (= g_{ij}dx^{i}dx^{j})$ for Kerr geometry as given by equation (3). From the analogy given by Apsel in equations (29-31) and using  the equation of action (as given by equation (28)), we can write the modified action equation as:
\begin{equation}
\tilde S=-\sum mc \int \sqrt{g_{tt}c^{2}dt^{2}+ g_{rr} dr^{2} + g_{\theta\theta} d\theta^{2} +g_{\phi\phi} d\phi^{2} + 2 g_{t \phi} c dt d\phi}-\frac{1}{c}\int  \Phi \ \rho \ d\Gamma  +\frac{1}{8\pi \ c} \int (H^{2}+E^{2}) \ d\Gamma
\end{equation}
 Finally, if we consider a rotating star having intense magnetic field (such as pulsars), then we can take an approximation $\Phi =0$ and E = 0. Now earlier obtained equation of action (32) can be rewritten as:
\begin{equation}
\tilde S=-\sum mc \int \sqrt{g_{tt}c^{2}dt^{2}+ g_{rr} dr^{2} + g_{\theta\theta} d\theta^{2} +g_{\phi\phi} d\phi^{2} + 2 g_{t \phi} c dt d\phi}  +\frac{1}{8\pi c}\int H^{2} d\Gamma
\end{equation}
From the above equation (33), we can write new form of $d \tilde s(= c \ d\tilde\tau)$ as:
 \begin{equation}
d \tilde s = \sqrt{g_{tt}c^{2}dt^{2}+ g_{rr} dr^{2} + g_{\theta\theta} d\theta^{2} +g_{\phi\phi} d\phi^{2} + 2 g_{t \phi} c dt d\phi}+ \frac{H^{2}dV c dt}{8\pi mc^{2}}
\end{equation}
For a sphere, the photon is emitted at a location on its surface where $dr = d\theta =0$, when the sphere rotates. Now above equation (34) can be written as:
\begin{equation}
c \ d\tilde\tau = cdt [\sqrt{{g_{tt}+ g_{\phi\phi}(\frac{d\phi}{c dt})^{2}} +2 g_{t \phi}(\frac{d\phi}{c dt})}+ \frac{H^{2} dV }{8\pi mc^{2}}]
\end{equation}
 above  equation (35) can be rewritten as:
\begin{equation}
\frac{d\tilde\tau}{dt} =  \sqrt{{g_{tt}+ g_{\phi\phi}(\frac{d\phi}{c dt})^{2}} +2 g_{t \phi}(\frac{d\phi}{c dt})}+ \frac{H^{2}
 }{8\pi \varrho_{m}c^{2}}
\end{equation}
where, $\varrho_{m} (= m/dV)$ is the scalar mass density. $\tilde\tau$ and t are proper time and world time respectively.\\
The quantity $\frac{d\phi}{c dt}$ is termed as angular velocity of frame dragging (as discussed in details \cite{Dubey and Sen 2014, Dubey and Sen 2015}).\\
In General relativity, redshift (Z) and redshift factor ($\Re$) are defined as \cite{Dubey and Sen 2014, Dubey and Sen 2015}:
\begin{equation}
\frac{d\tilde\tau}{dt}=\frac{\omega_{ob}}{\omega_{em}} =\Re = \frac{1}{Z+1}=  \frac{\lambda_{em}}{\lambda_{ob}}
\end{equation}
where $\omega$ and $\lambda$ denote frequency and wavelength, respectively. Emitter's and observer's frame of reference are indicated by subscripts \textit{em}  and \textit{ob}. A redshift (Z) of zero corresponds to an un-shifted line, whereas $Z<0$ indicates blue-shifted emission and $Z>0$ red-shifted emission. A redshift factor ($\Re$) of unity corresponds to an un-shifted line, whereas $\Re<1$ indicates red-shifted emission and $\Re>1$ blue-shifted emission.
\\ From equation (36) and (37), we can write redshift factor $\Re$ as:
\begin{equation}
\Re=\sqrt{{g_{tt}+ g_{\phi\phi}(\frac{d\phi}{c dt})^{2}} +2 g_{t \phi}(\frac{d\phi}{c dt})}+ \frac{ H^{2} }{8\pi \varrho_{m}c^{2}}
\end{equation}
The above expression (38) is the expression of gravitational redshift from rotating body having intense magnetic field.  The first term of the expression is due to mass and rotation effect, which is gravitational redshift from rotating body and given as:
\begin{equation}
\Re_{mass+rotation}= \sqrt{{g_{tt}+ g_{\phi\phi}(\frac{d\phi}{c dt})^{2}} +2 g_{t \phi}(\frac{d\phi}{c dt})}
\end{equation}
The second term is due to the presence of intense magnetic field in the rotating body, which is  an additional magnetic redshift and given as:
\begin{equation}
\Re_{mag}(H,\varrho_{m})= \frac{H^{2} }{8\pi \varrho_{m}c^{2}}
\end{equation}
   For a typical neutron star the value of magnetic field is $H \sim 10^{12}$ Gauss and density is $\varrho_{m}\sim 10^{15}-10^{16} \ \frac{gm}{c^{3}}$  (page 293 of Straumann \cite{Straumann}). From equation (40), we can obtain an additional magnetic redshift by using the values of scalar mass density ($\varrho_{m}$) and magnetic field (H).
\section{\label{sec:Concl}Conclusions}

In the present work the expression of gravitational redshift from rotating body having intense magnetic field has been derived by using the action function of the electromagnetic fields. The first term of the derived expression is due to mass and rotation effect, which is gravitational redshift
from rotating body. While the second term is due to the presence of intense magnetic field in the rotating body, which is an additional magnetic redshift. If we ignore the  electric field (E) and magnetic field (H) contribution then we can obtain the corresponding expression for redshift factor ($\Re$) in Kerr Geometry \cite{Dubey and Sen 2014}.

\end{document}